\documentclass{svproc}
\usepackage{url}

\usepackage{epsf,color,colordvi}
\usepackage{caption}
\usepackage{graphicx}
\usepackage{amsmath}
\usepackage{times}
\usepackage{subfig}
\usepackage{float}
\usepackage{ulem}  % This package redefines \emph{} to underline
\usepackage{placeins}
\usepackage{float}

\newcommand{\ba}{\begin{array}}
	\newcommand{\ea}{\end{array}}
\def\beq{\begin{equation}}
\def\eeq{\end{equation}}
\def\bea{\begin{eqnarray}}
\def\eea{\end{eqnarray}}
\def\nn{\nonumber}

\def\roughly#1{\mathrel{\raise.3ex\hbox
		{$#1$\kern-.75em\lower1ex\hbox{$\sim$}}}}

\def\sla#1{\raise.15ex\hbox{$/$}\kern-.57em #1}% Feynman slash

\def\bd{B_d^0}

\def\order{\lower 1.8ex \hbox{\LARGE\~{}}}

\def\bd0tau{B\to D \tau\nu_{\tau}}

\def\be {\begin{equation}}
\def\ee {\end{equation}}

\begin{document}
\mainmatter              % start of a contribution
\title{$b\to cl\nu$ anomalies in light of vector and scalar
interactions}
\titlerunning{$b\to cl\nu$ anomalies}  % abbreviated title (for running head)
\author{Aritra Biswas\inst{1} 
%\and M V N Murthy\inst{1} \and Nita Sinha\inst{1}
% \and Roger Temam\inst{2}
% Jeffrey Dean \and David Grove \and Craig Chambers \and Kim~B.~Bruce \and
% Elsa Bertino
}
%
%\authorrunning{Aritra Biswas et al.} % abbreviated author list (for running head)
%\tocauthor{Aritra Biswas, 
% Roger Temam, Jeffrey Dean, David Grove,
% Craig Chambers, Kim B. Bruce, and Elisa Bertino
%}
%
\institute{Indian Association for the cultivation of Science\\
\email{tpab2@iacs.res.in},\\ home page:
\texttt{http://inspirehep.net/author/profile/Arita.Biswas.1}
% \and
% Universit\'{e} de Paris-Sud,
% Laboratoire d'Analyse Num\'{e}rique, B\^{a}timent 425,\\
% F-91405 Orsay Cedex, France
}

\maketitle              % typeset the title of the contribution

\begin{abstract}
We perform a model independent analysis of the charged current $b\to cl\nu$ anomalies under the presence of scalar and vector interactions. The analysis is carried out in two stages: (a) under the presence of both (left-handed) vector and scalar interactions and (b) under the presence of scalar interactions alone. We find that even after stringent bounds from similar quark-level processes such as $B_c\to\tau\nu$, such scenarios have the potential to explain the aforementioned anomalies. Contrary to the general notion, we show that even scalar interactions alone can explain such anomalies, provided they are complex. However, extended scalar sector models are unable to comply with these anomalies to $\sim 3\sigma$. We further illustrate our results with the help of three benchmark models corresponding to the presence of (i) both scalar and vector (ii)  real scalar and (iii) complex scalar interactions.
% We would like to encourage you to list your keywords within
% the abstract section using the \keywords{...} command.
\end{abstract}
\section{Introduction}
Over the past few years, there have been constant and consistent reports from experimental collaborations such as LHCb, Belle and BaBar about flavour observables with deviations of more than $3\sigma$ in exclusive $B\to D^{*}$~\cite{Lees:2012xj,Lees:2013uzd,Huschle:2015rga,Sato:2016svk,Aaij:2015yra,Hirose:2016wfn,Hirose:2017dxl,Aaij:2017uff,Aaij:2017deq} and $B\to J/\psi$ transitions. Both of these exclusive processes have the underlying sub-quark transition $b\to cl\nu$. These results are believed to be the hints of lepton-flavour universality violating (LFUV) type new physics (NP). We investigate the prospect of scalar and vector type NP's in explaining such deviations. We initially work from a model independent perspective and then illustrate our results further using the models: (i) Non-minimal universal extra dimensions (NMUED) for the case with one scalar and one vector NP operator, (ii) Goergi-Michacek (GM) model for the case of a single scalar NP operator preceded by a real Wilson coefficient (WC) and (iii) Leptoquark (LQ) model for the a single scalar NP operator preceded by a complex WC.
\section{Current Status: Theory and Experiment}
The present global average for the  $\mathcal{R}(D^{(*)})$ anomalies are about $4\sigma$ away from the corresponding SM results. Fig.~\ref{fig:excGM} and table.~\ref{tab:RDRDsPtau} summarize the current theoretical and experimental status for these anomalies. The SM average is the arithmetic mean of the results from~\cite{Bigi:2016mdz,Bernlochner:2017jka,Bigi:2017jbd,Jaiswal:2017rve}.
	\begin{table}
	\small
%	\ht
%		\begin{center}
% 			\begin{ruledtabular}
				\begin{tabular}{cccccc}
					& $\mathcal{R}(D)$  & $\mathcal{R}(D^*)$  &	Correlation	& $P_{\tau}(D^*)$ &$\mathcal{R}(J/\psi)$\\
					\hline
					%			\noalign{\vskip1pt}
					%			SM  		   	& $0.300(8)$ \cite{Na:2015kha}   	& $0.252(3)$ \cite{Kamenik:2008tj} 		&&\\
					%			& $0.299(11)$ \cite{Lattice:2015rga}	& 						&&\\
					%			& $0.299(3)$ \cite{Bigi:2016mdz}	& $0.262(10)$ \cite{Bigi:2017jbd}		&&\\
					%			& $0.299(3)$				& $0.257(3)$ 		& $0.44$ \cite{Bernlochner:2017jka}&\\
					SM			& $0.299(3)$		& $0.258(6)$		  &						&  $-0.491(25)$& $0.249(42)$(LFCQ)\\
                                            			        &          		&        		  &						&              & $0.289(28)$(PQCD)\\
					\hline
 					Babar~   	& $0.440(58)_{st.}(42)_{sy.} $ 	& $0.332(24)_{st.}(18)_{sy.}$ & $-0.27$ & &\\
 					Belle (2015)& $0.375(64)_{st.}(26)_{sy.}$ 	& $0.293(38)_{st.}(15)_{sy.}$ & $-0.49$&&\\
 					Belle (2016)-I & -			 	& $0.302(30)_{st.}(11)_{sy.}$  & & &\\
 					Belle (2016)-II & - 			& $0.270(35)_{st.}~^{+ 0.028}_{-0.025}$ & 0.33 & $ -0.38(51)_{st.}~^{+0.21}_{-0.16}$ &\\
 					%			Full Dataset) 		& & & \\
 					LHCb (2015) & - 				& $0.336(27)_{st.}(30)_{sy.}$  & & &\\
 					LHCb (2017) & - 				& $0.286(19)_{st.}(25)_{sy.}(21)$  & & &\\
 					\hline
 					World Avg.	& $0.407(39)_{st.}(24)_{sy.}$	& $0.304(13)_{st.}(7)_{sy.}$ & $0.20$  & &$0.71(17)_{st.}(18)_{sy.}$\\
				\end{tabular}
 				\caption{\small Present status (both theoretical and experimental) of $\mathcal{R}(D)$, $\mathcal{R}(D^*)$ and $P_{\tau}(D^*)$. First uncertainty is statistical and the second one is systematic. The first row lists the arithmetic mean for the SM calculations reported in HFLAV.} 
				\label{tab:RDRDsPtau}
% 			\end{ruledtabular}
%		\end{center}
	\end{table}

%  	\begin{table}
\begin{figure}
		\centering
		\includegraphics[scale=0.3]{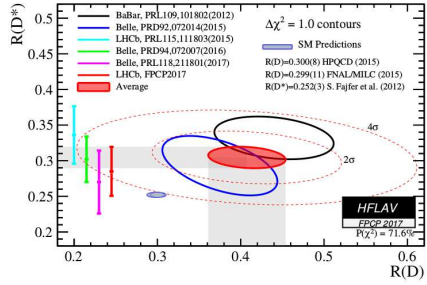}
  		\caption{\small Global average for $\mathcal{R}(D)$, $\mathcal{R}(D^*)$ and $P_{\tau}(D^*)$ and the deviation from the SM result.}
		\label{fig:excGM}
\end{figure}
\section{Formalism}
	The most general effective Hamiltonian describing the $b\to c\tau \nu$ transitions, with %a new scalar type four-fermi operator ${\cal O}_{S_1}$ in the lowest dimension,
	all possible four-fermi operators in the lowest dimension (with left-handed neutrinos)
	is given by:
	\beq\label{eq1}
	{\cal H}_{eff} = \frac{4 G_F}{\sqrt{2}} V_{cb}\Big[( 1 + C^\ell_{V_1}) {\cal O}_{V_1} +
	C^\ell_{V_2} {\cal O}_{V_2} + C^\ell_{S_1} {\cal O}_{S_1}+ C^\ell_{S_2} {\cal O}_{S_2}+ C^\ell_{T}{\cal O}_{T}\Big],
	\eeq
	
%	\begin{equation}
%	{\cal H}_{eff} = \frac{4 G_F}{\sqrt{2}} V_{cb}\Big[ {\cal O}_{V} + C_{S}^{\ell} {\cal O}_{S_1} \Big],
%	\label{eq1}
%	\end{equation}
%	with ${\cal O}_{V} = ({\bar c}_L \gamma^\mu b_L)({\bar \tau}_L \gamma_\mu \nu_{L})$ and ${\cal O}_{S_1} = ({\bar c}_L  b_R)({\bar \tau}_R \nu_{L})$
	where the operator basis is defined as
	\bea
	{\cal O}_{V_1} &=& ({\bar c}_L \gamma^\mu b_L)({\bar \tau}_L \gamma_\mu \nu_{\tau L}) \nn, \\
	{\cal O}_{V_2} &=& ({\bar c}_R \gamma^\mu b_R)({\bar \tau}_L \gamma_\mu \nu_{\tau L}) \nn, \\
	{\cal O}_{S_1} &=& ({\bar c}_L  b_R)({\bar \tau}_R \nu_{\tau L}) \nn, \\
	{\cal O}_{S_2} &=& ({\bar c}_R b_L)({\bar \tau}_R \nu_{\tau L}) \nn, \\
	{\cal O}_{T}   &=& ({\bar c}_R \sigma^{\mu\nu} b_L)({\bar \tau}_R \sigma_{\mu\nu} \nu_{\tau L}),
	\label{eq2}
	\eea
	and the corresponding Wilson coefficients are given by $C_X (X=V_1,V_2,S_1,S_2,T)$. 
	%In our case, $C_{S_1}\equiv C_{S_{\ell}}$. Rest of the coefficients are all zero.
	We are interested in the new scalar interaction ${\cal O}_{V_1}$ and ${\cal O}_{S_1}$, and thus we turn all other Wilson Coefficients to zero for this analysis.
	
	Subject to the above hamiltonian, one can construct the differential decay rate for a particular exclusive decay, involving the NP WC's, the CKM elements and the corresponding hadronic form factors. The measurable observables are ratios fo these integrated decay rates with different leptons in the final states. The ratio cancels uncertainties due to the CKM elements completely, and also those due to the form factors to a large extent. For the theoretical details regarding the obserbables, the corresponding form factors and the constraints, the interested reader can look into~\cite{Biswas:2017vhc,Biswas:2018jun} and the references therein.  
\section{Analysis}
\begin{figure}
\tiny
		\subfloat{\includegraphics[width=0.25\textwidth]{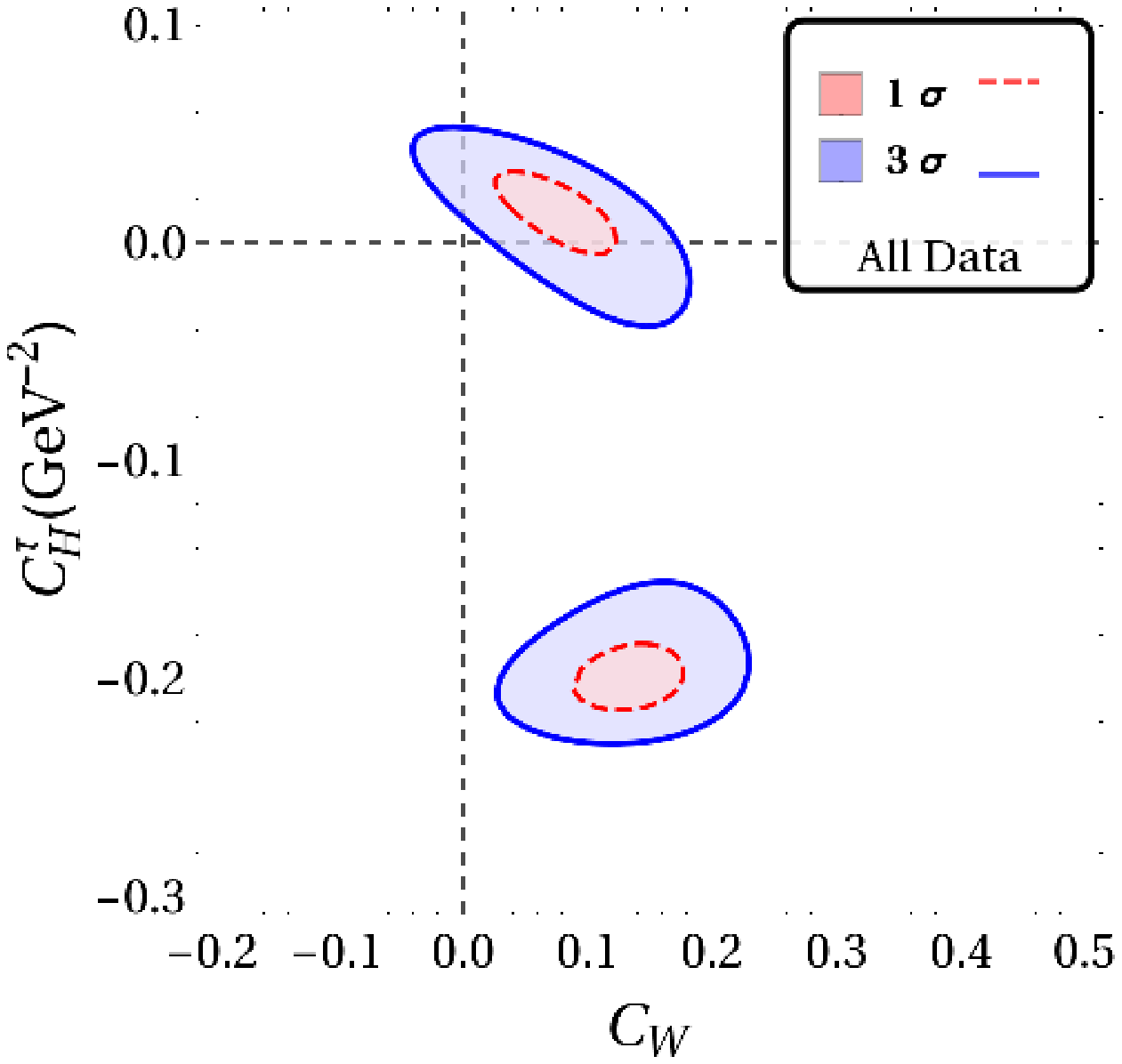}}
		\subfloat{\includegraphics[width=0.25\textwidth]{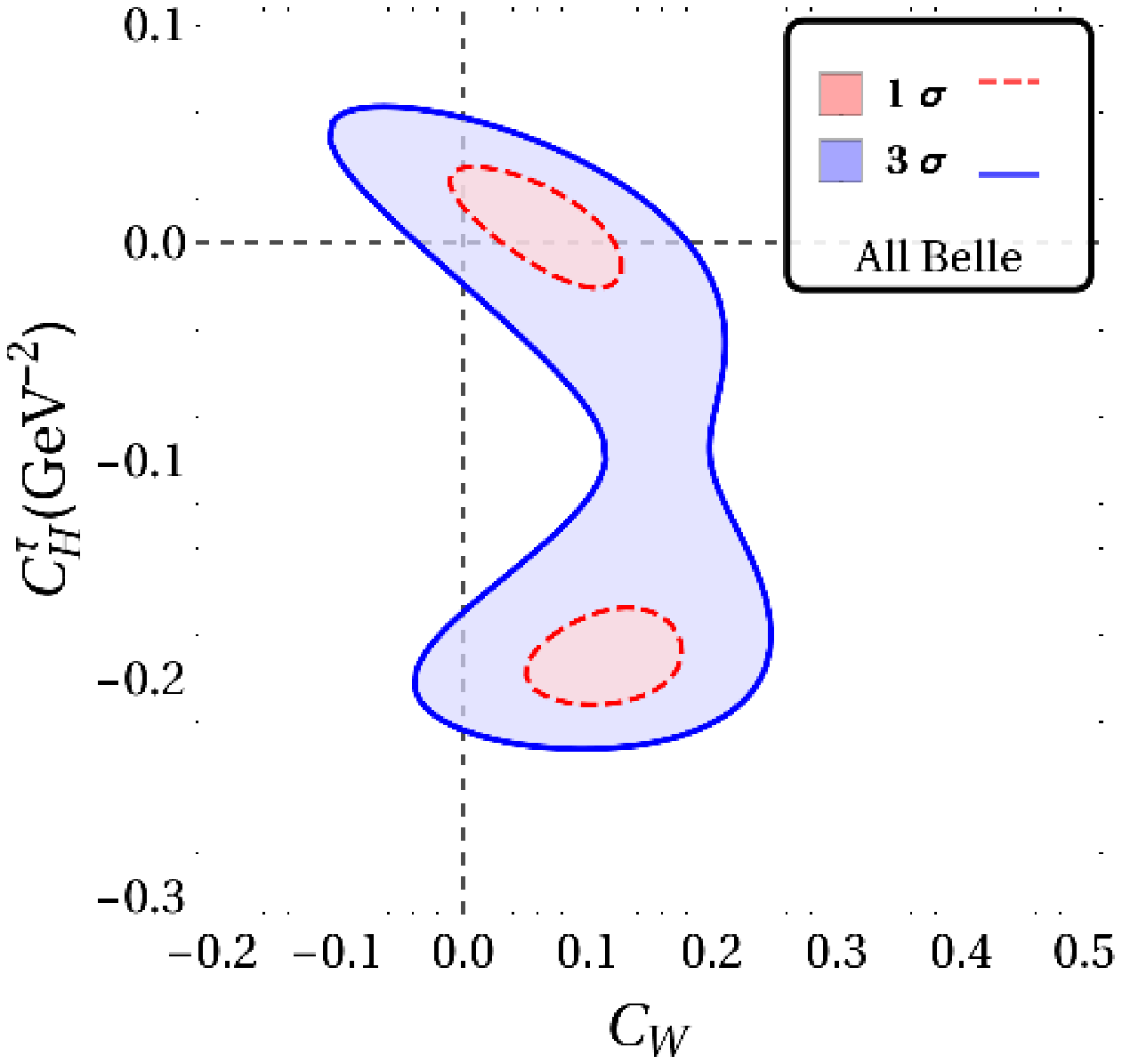}}
		\subfloat{\includegraphics[width=0.25\textwidth]{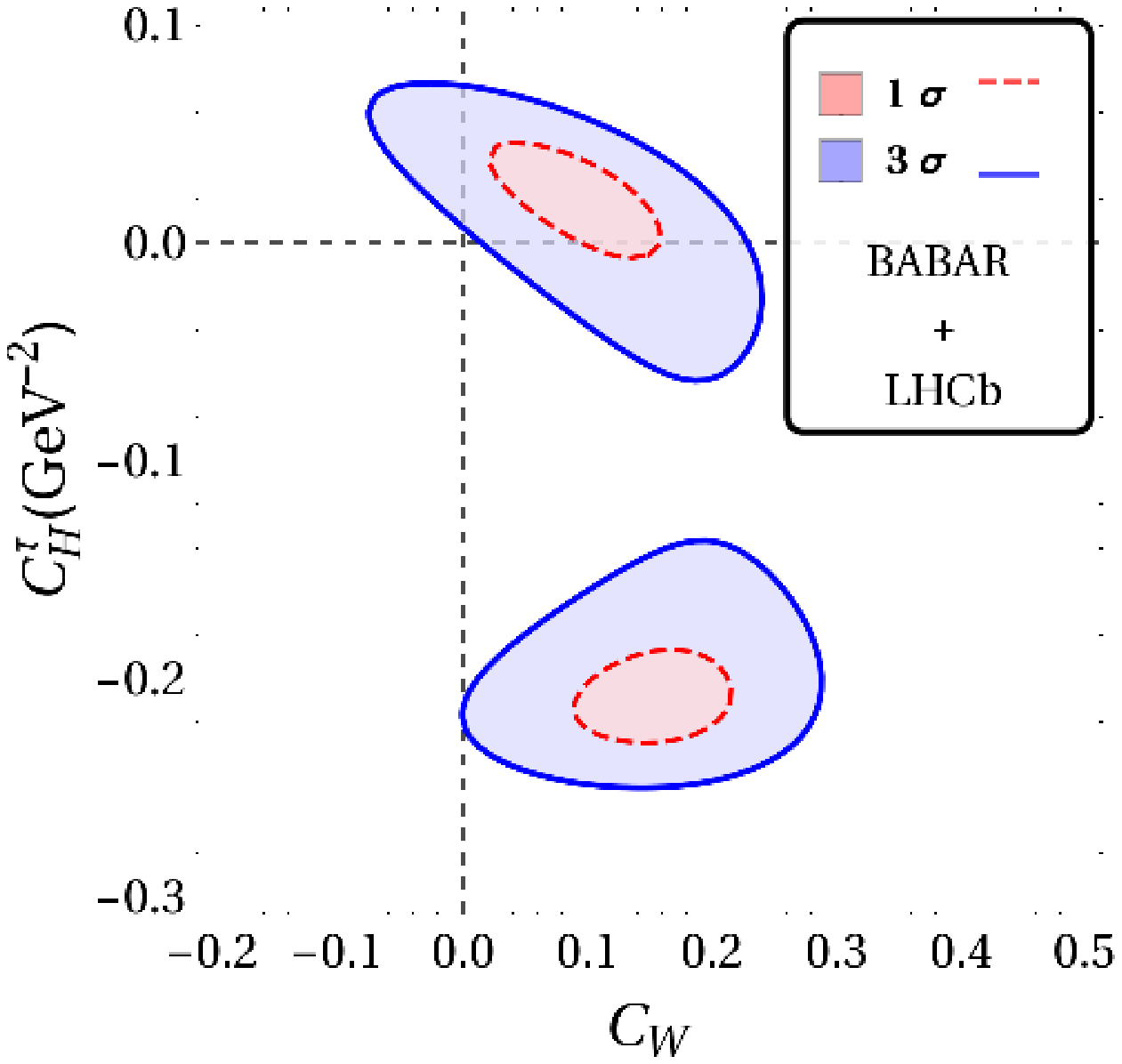}}
		\subfloat{\includegraphics[width=0.25\textwidth]{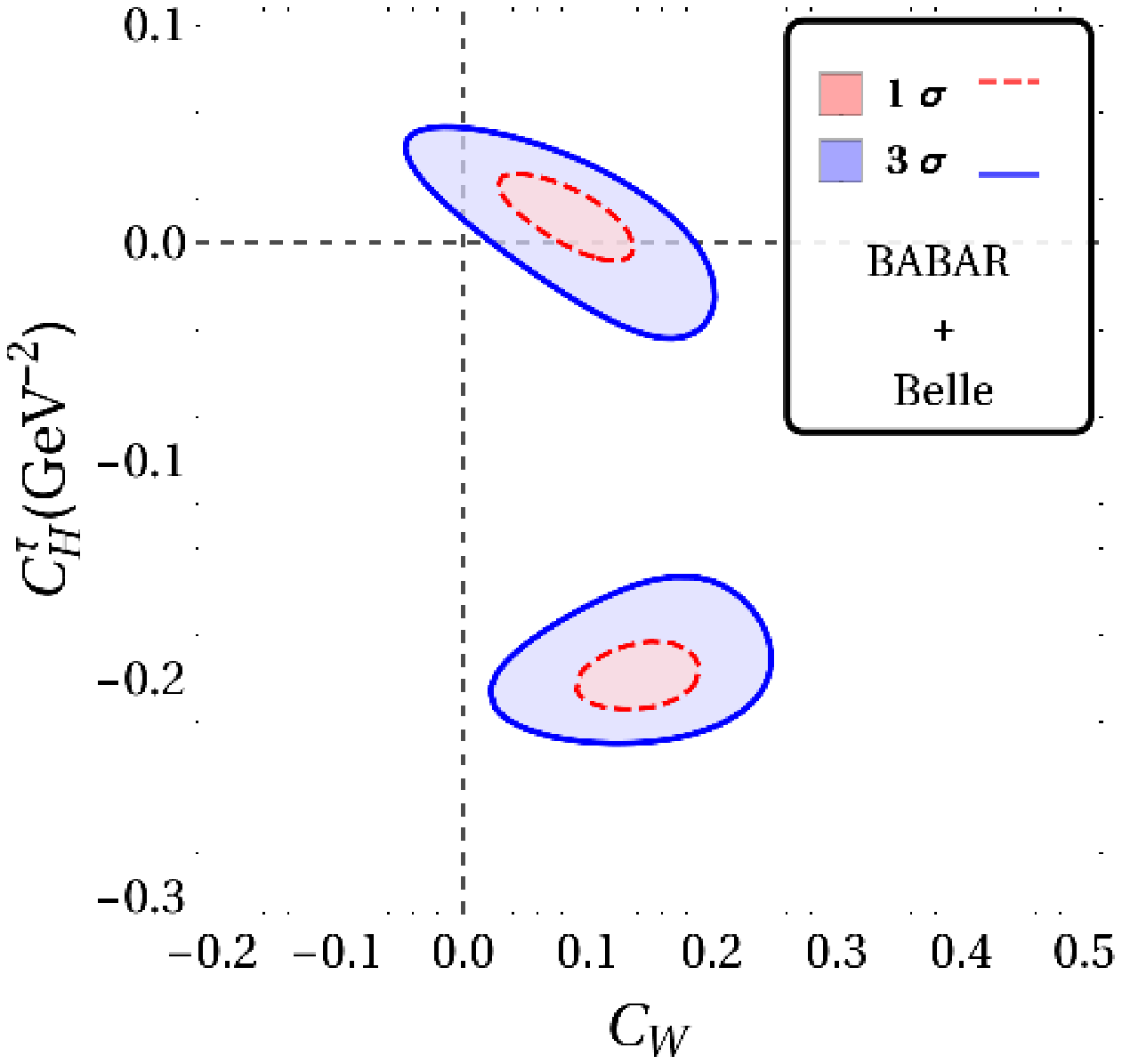}}\\
		\subfloat{\includegraphics[width=0.25\textwidth]{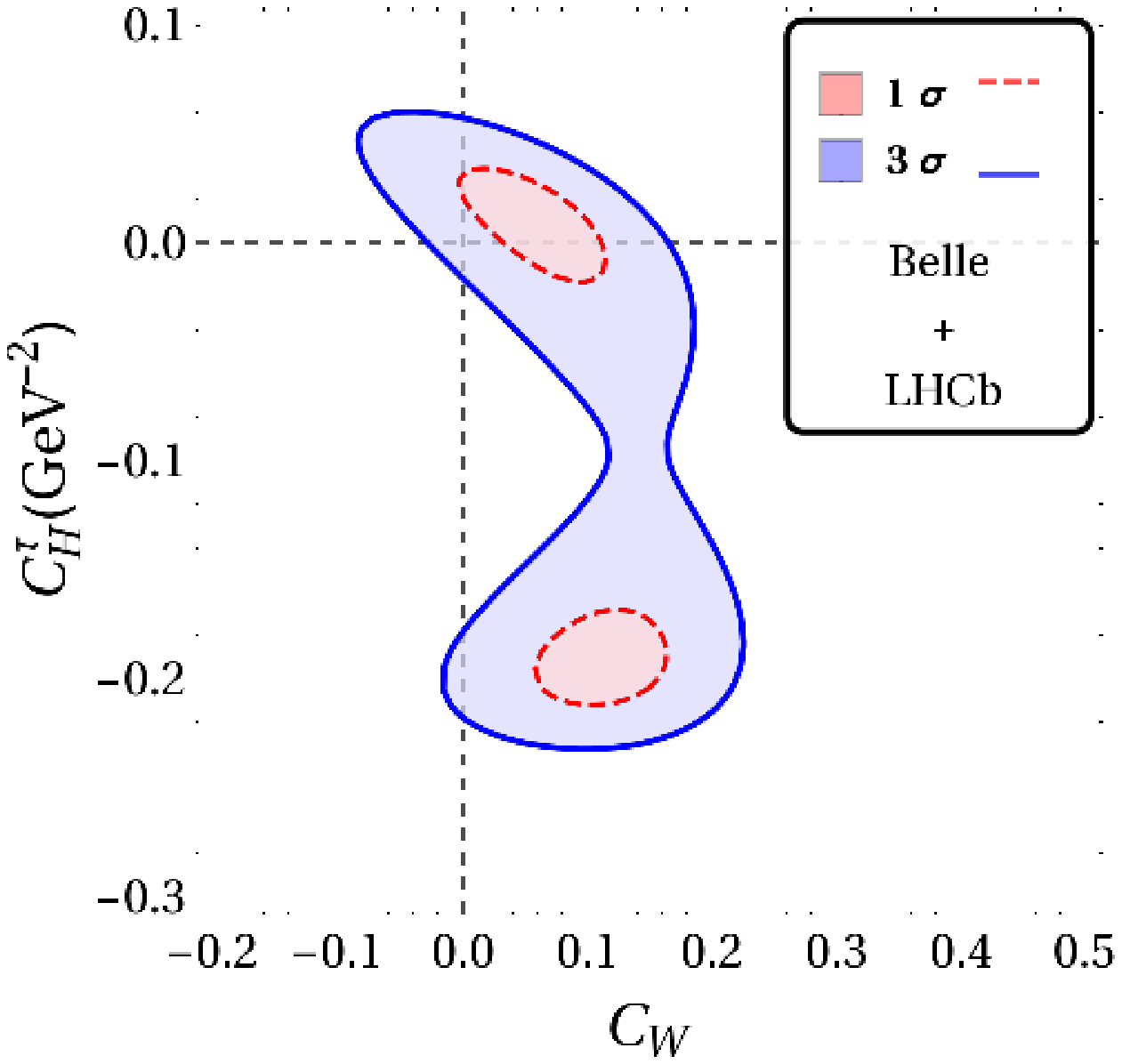}}
		\subfloat{\includegraphics[width=0.25\textwidth]{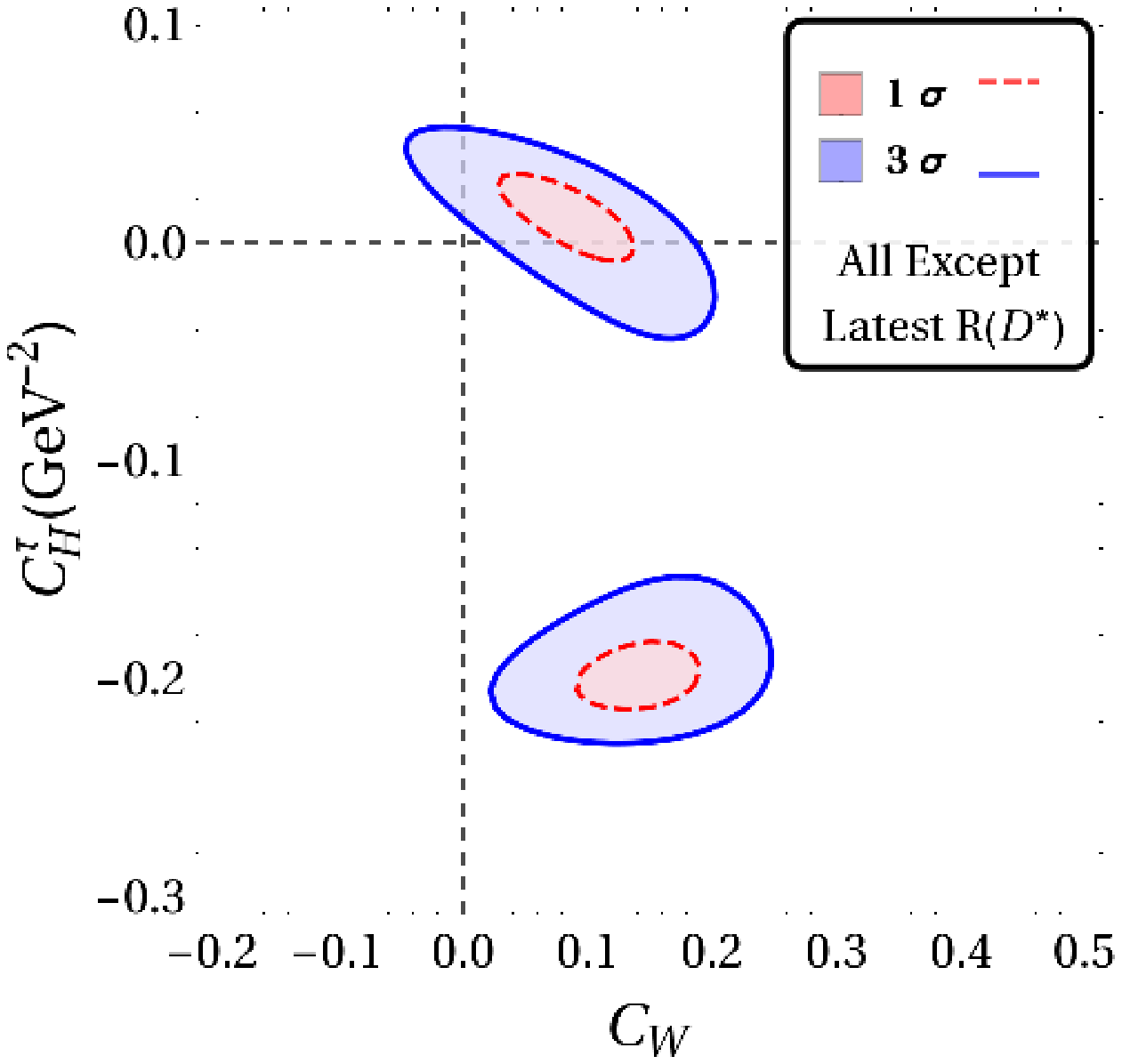}}
		\subfloat{\includegraphics[width=0.25\textwidth]{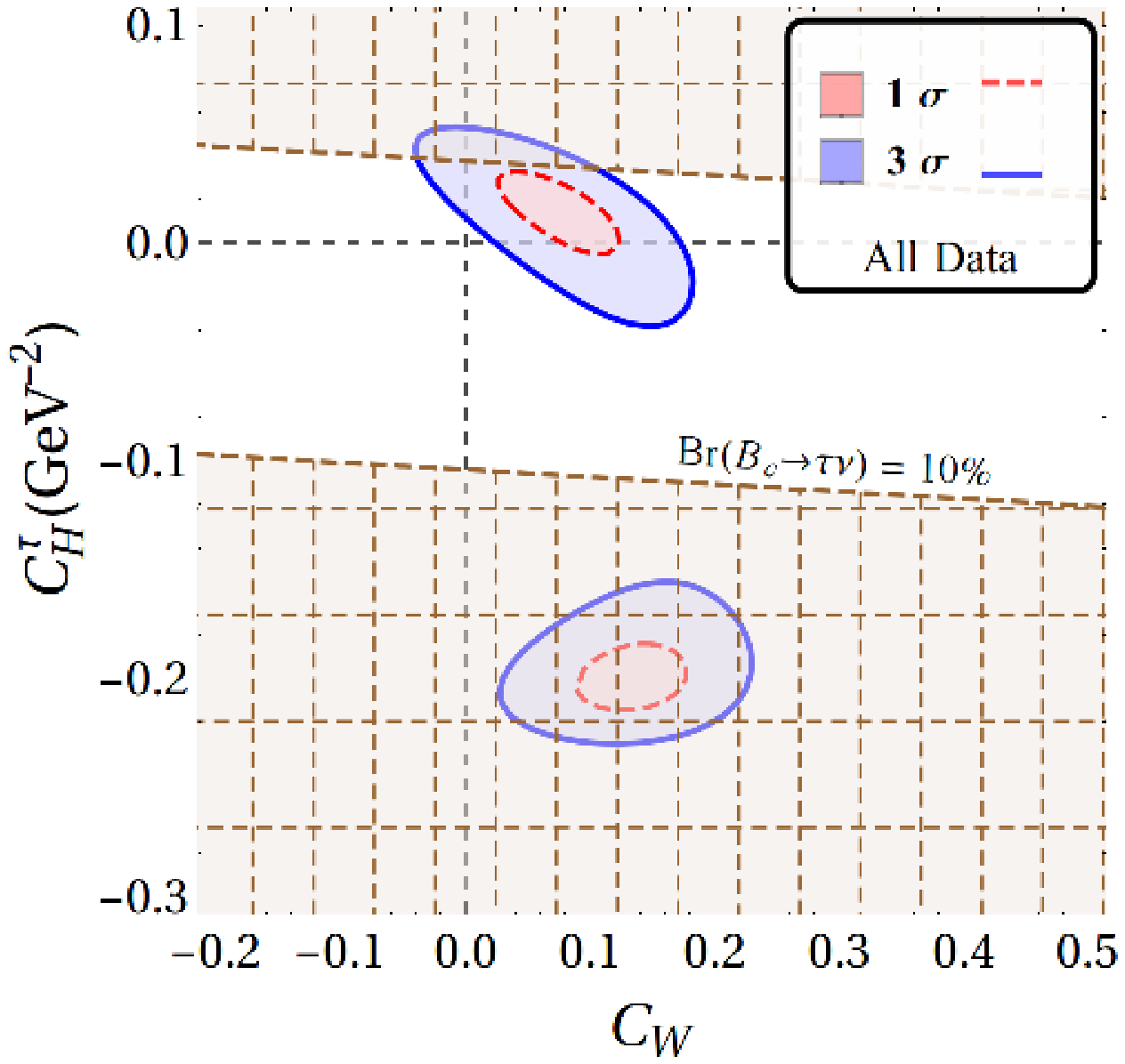}}
		\subfloat{\includegraphics[width=0.25\textwidth]{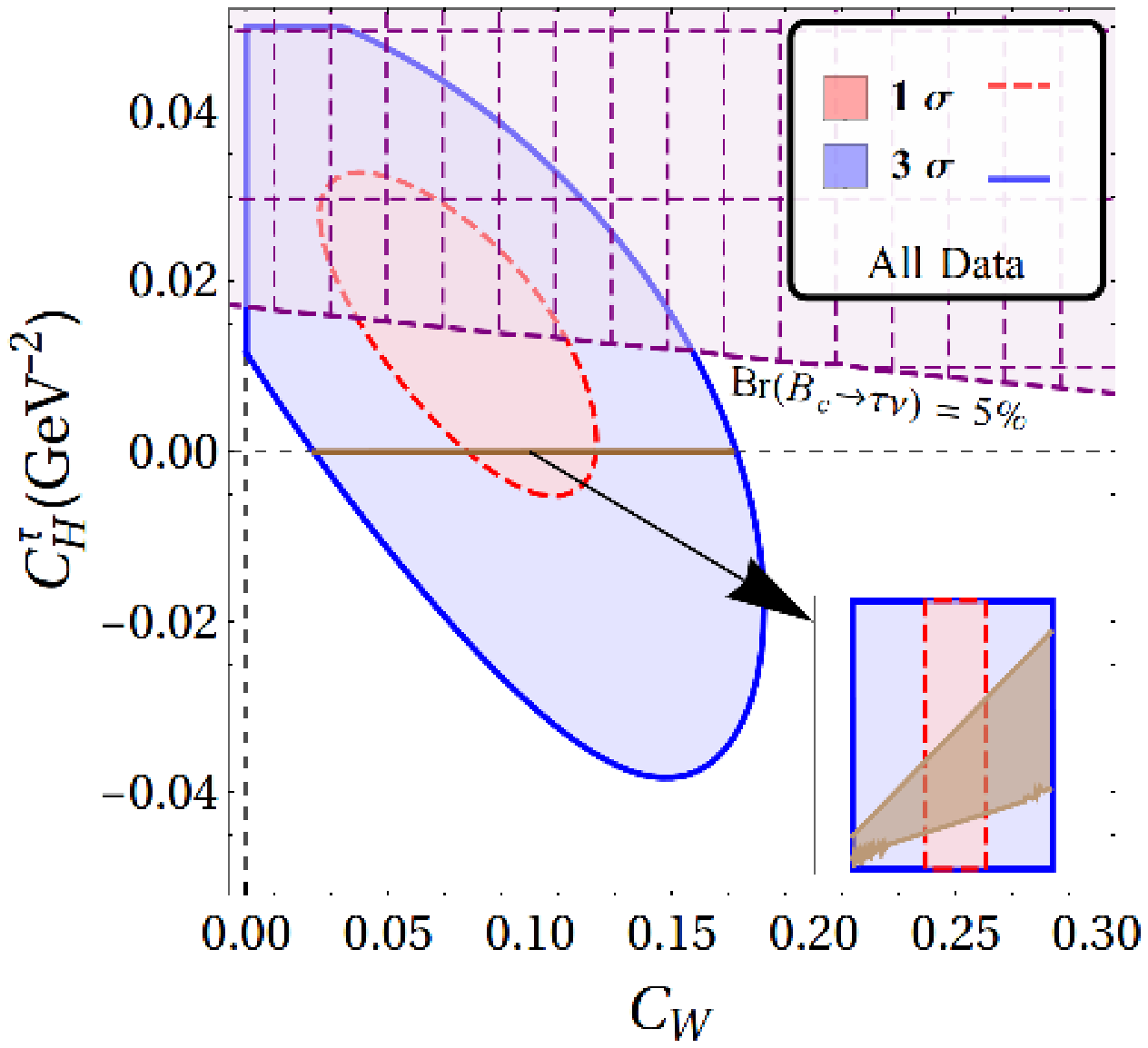}}
 \caption{\small $\mathcal{R}(D^{(*)})$ fit results corresponding to separate fits listed in table \ref{tab:res1} for the case with both $O_{V_1}$ and $O_{S_1}$. Red(dotted) and blue(solid) lines enclose $1 \sigma$ ($\Delta\chi^2 = 2.30$) and $3 \sigma$ ($\Delta\chi^2 = 11.83$) regions respectively. Only the gridlines corresponding to $C_W=C_{V_1}$ and $C^{\tau}_H = 0$ are shown, such that there intersection point represents SM. The hatched regions in the last two figures show the constraints coming from $B_c\to\tau\nu$. }
\label{fig:chiplts}
\end{figure}

The results for our fits with a single vector and scalar type NP are displayed in fig.~\ref{fig:chiplts} and table.~\ref{tab:res1}. In what follows $C_{S_1}=-C_H ~m_b ~m_{\ell}$. The WC's are considered to be real. It is clear that for all combinations of results shown in fig.~\ref{fig:chiplts}, there is a two-fold ambiguity in the best-fit results. One of these points is closer to SM than the other and this is the one that is important in constraining NMUED. We also note that while the results from Belle and LHCb are consistent with SM within $3 \sigma$, for any and all other combination of results, the SM is away from the best fit point by more than $3 \sigma$ in the $C_W$ - $C^{\tau}_{H}$ plane.  
\begin{table}
% 		\centering
		\small
			%\begin{ruledtabular}
%                      \resizebox{18cm}{!}{
				\begin{tabular}{|c|c|c|c|c|c|c|c|c|c|c|}\hline
					& \multicolumn{2}{|c|}{Without $\mathcal{R}_{J/\psi}$} & \multicolumn{4}{|c|}{With $\mathcal{R}_{J/\psi}$} & \multicolumn{2}{|c|}{Fit Results} & \multicolumn{2}{|c|} {Observable values}\\  
					\cline{2-11}
					& & &\multicolumn{2}{|c|}{PQCD} & \multicolumn{2}{|c|}{LFCQ} & & & &\\ 
					\cline{4-7}
					Datasets & $\chi^2_{min}$ & $p$-value & $\chi^2_{min}$ & $p$-value & $\chi^2_{min}$ & $p$-value & Re($C_H$) & Im($C_H$) & $\mathcal{R}(D^*)$ & $\mathcal{R}(D)$ \\
					& /DoF  & (\%) & /DoF  & (\%) & /DoF  & (\%) & (GeV$^{-2}$)  & (GeV$^{-2}$) & &\\
					\hline
					All Data		& 9.22/8	& 23.72		& 11.86/9	&	15.76	& 12.38/9	&	13.51	& -0.031(8)		& 0.000(73) & 0.2746(25)&0.448(42)\\
					Belle			& 1.71/4	& 63.54		& 4.39/5	&	35.63	& 4.89/5	&	29.83	& -0.023(11)	& 0.000(87)  & 0.2674(33)&0.406(60)\\
					Babar~+LHCb	& 6.42/3	& 4.03		& 9.00/4	&	2.92 	& 9.54/4	&	2.29	& -0.042(11)	& 0.000(84)  & 0.2764(34)&0.508(58)\\
					Babar~+ Belle	& 6.71/6	& 24.31		& 9.35/7	&	15.48	& 9.87/7	&	13.03	& -0.030(8)		& 0.000(74)  & 0.2724(25)&0.445(43)\\
					Belle + LHCb	& 4.70/6	& 45.41		& 7.37/7	&	28.82	& 7.88/7	&	24.72	& -0.025(11)	& 0.000(78)  & 0.2700(34)&0.414(59)\\
					All $\mathcal{R}_{D^{*}}$	& 2.37/5	& 66.78		& 4.31/6 	&	50.53	& 4.99/6	&	41.67	& -	& -  & &\\
					No $P_{\tau}(D^*)$	& 9.21/7	& 16.23	& 11.84/8	&	10.58	& 12.36/8	&	8.92	& -0.031(8)		& 0.000(72)  &0.2746(25) &0.448(42)\\ \hline
				\end{tabular}
				\caption{\small Results of fits for different combinations of experimental data-points along with our predictions for the charged current observables for the case with both $O_{V_1}$ and $O_{S_1}$. The last two columns are obtained from fits without treating $\mathcal{R}_{J/\psi}$ as a data point.} 
				\label{tab:res1}
			%\end{ruledtabular}
		%\end{center}
	\end{table}

\begin{figure}
		\centering
		\subfloat{\includegraphics[width=0.25\textwidth]{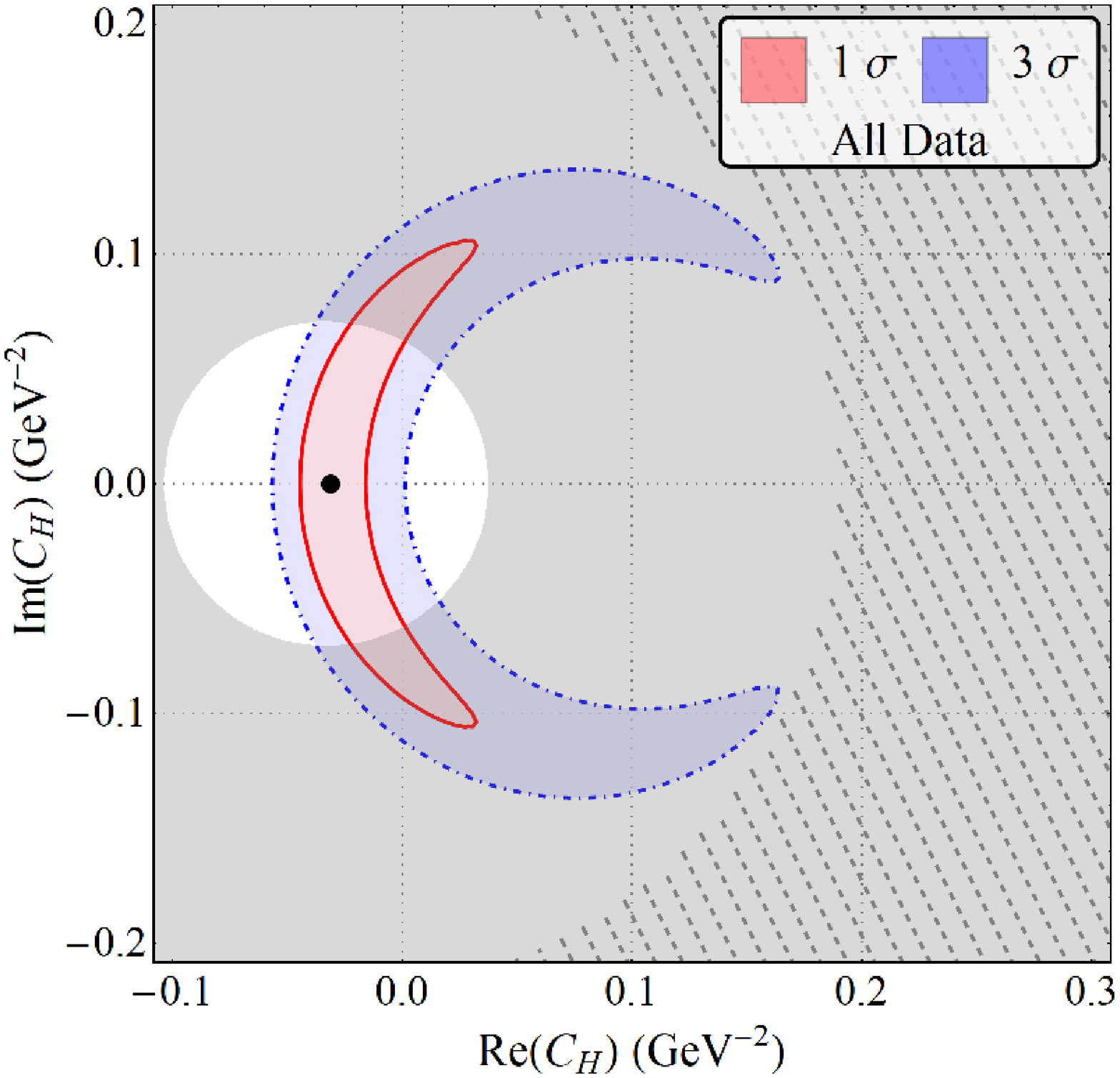}}
		\subfloat{\includegraphics[width=0.25\textwidth]{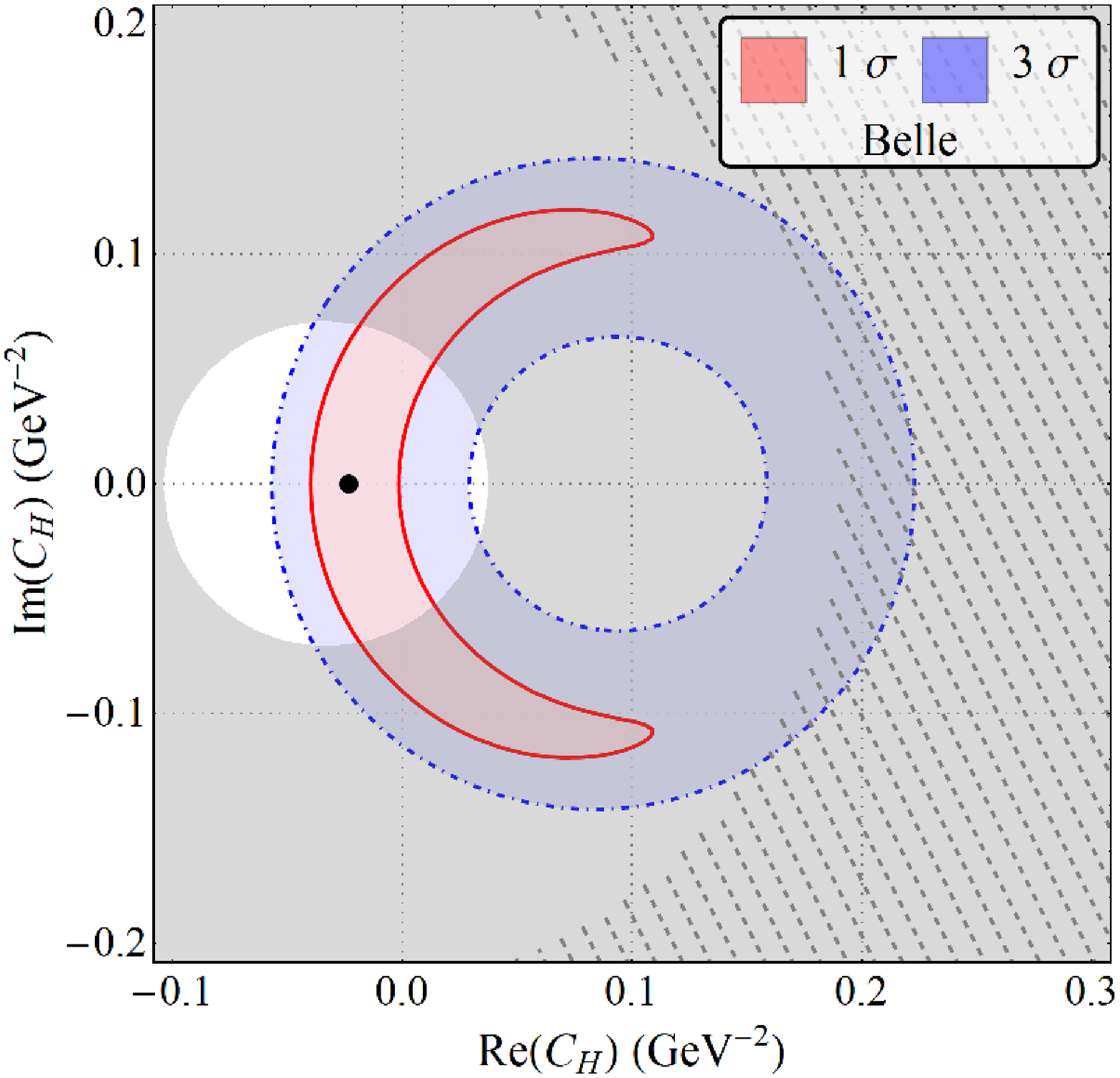}}
		\subfloat{\includegraphics[width=0.25\textwidth]{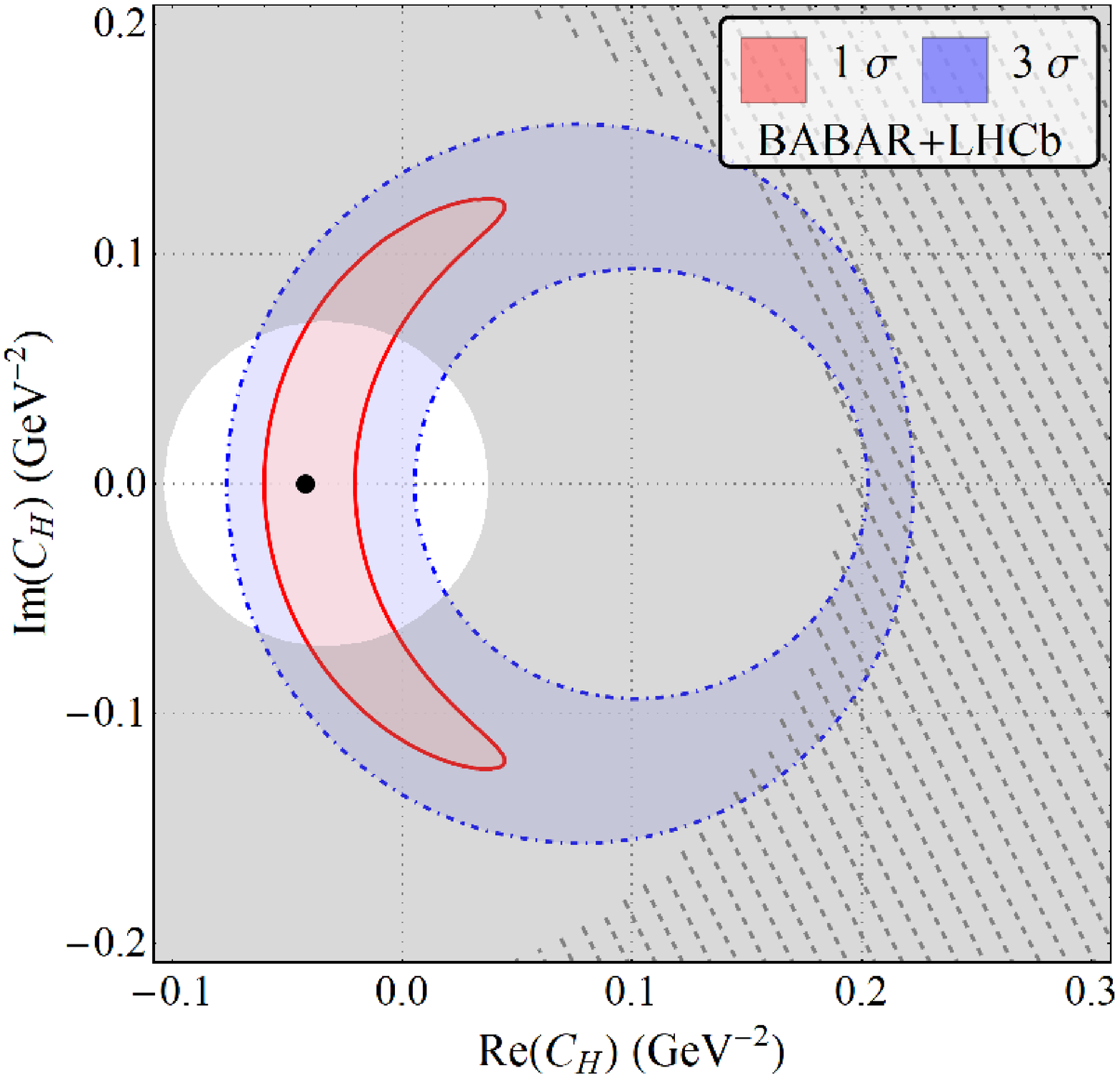}}\\
		\subfloat{\includegraphics[width=0.25\textwidth]{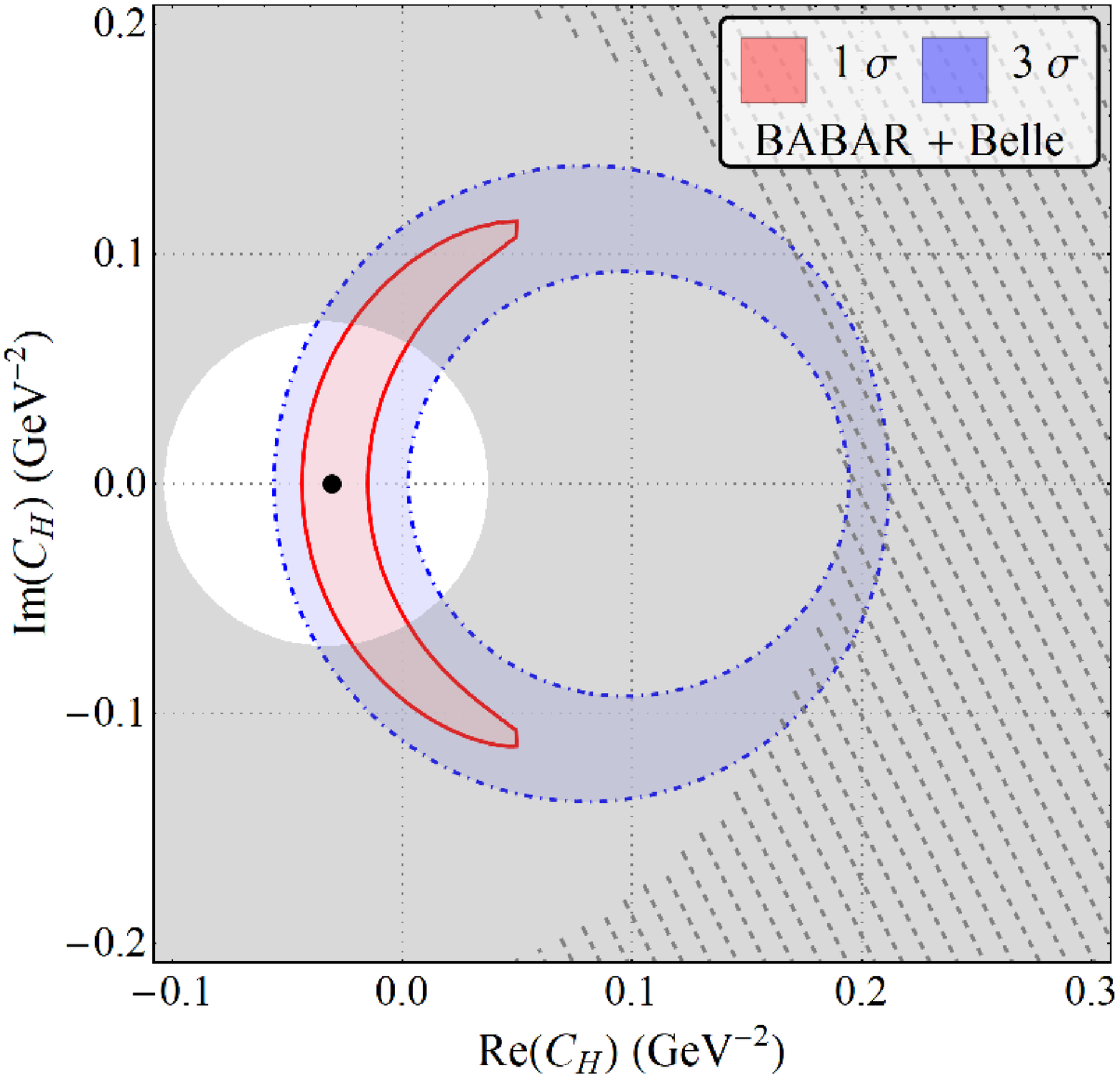}}
		\subfloat{\includegraphics[width=0.25\textwidth]{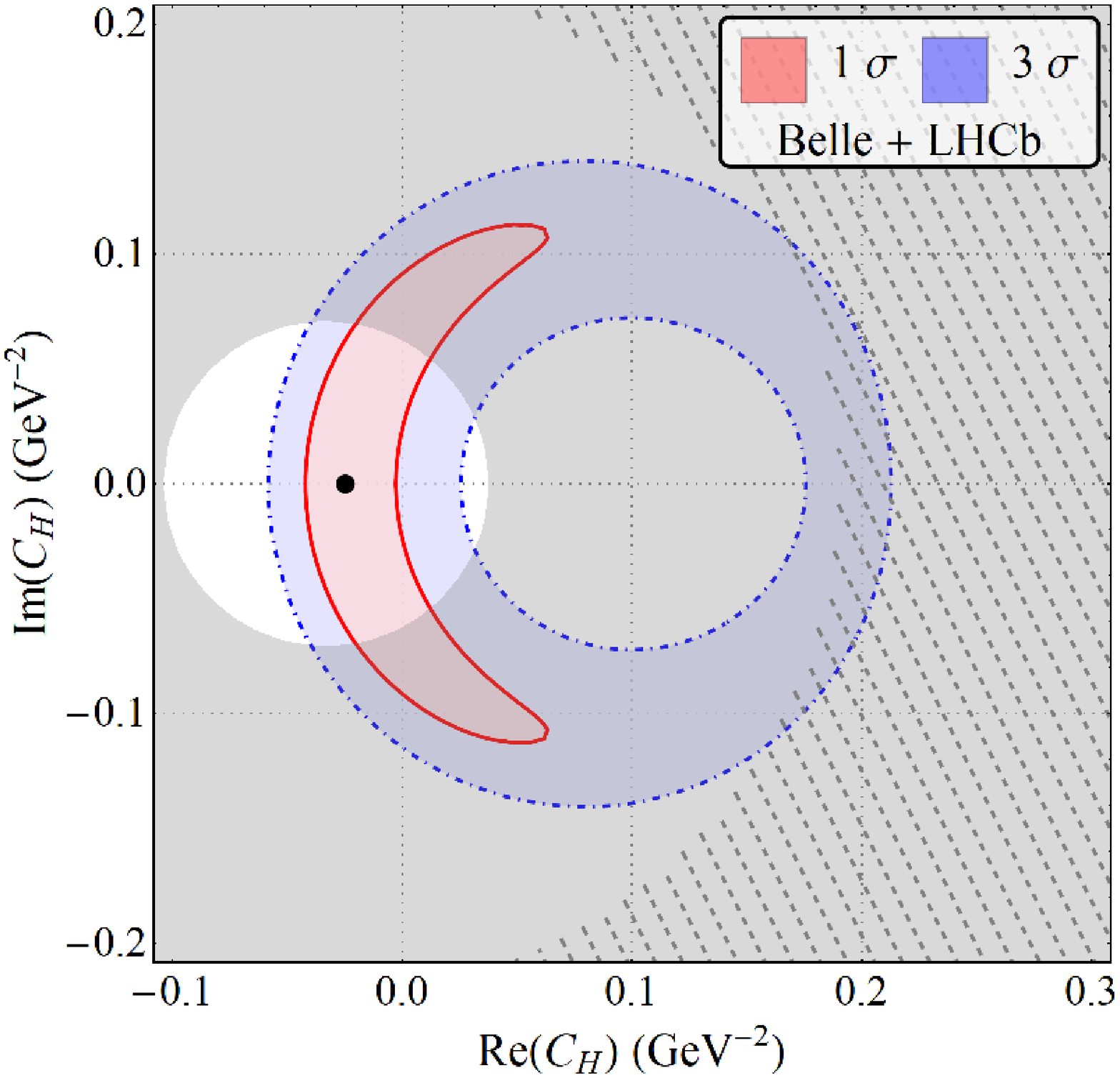}}
		\subfloat{\includegraphics[width=0.25\textwidth]{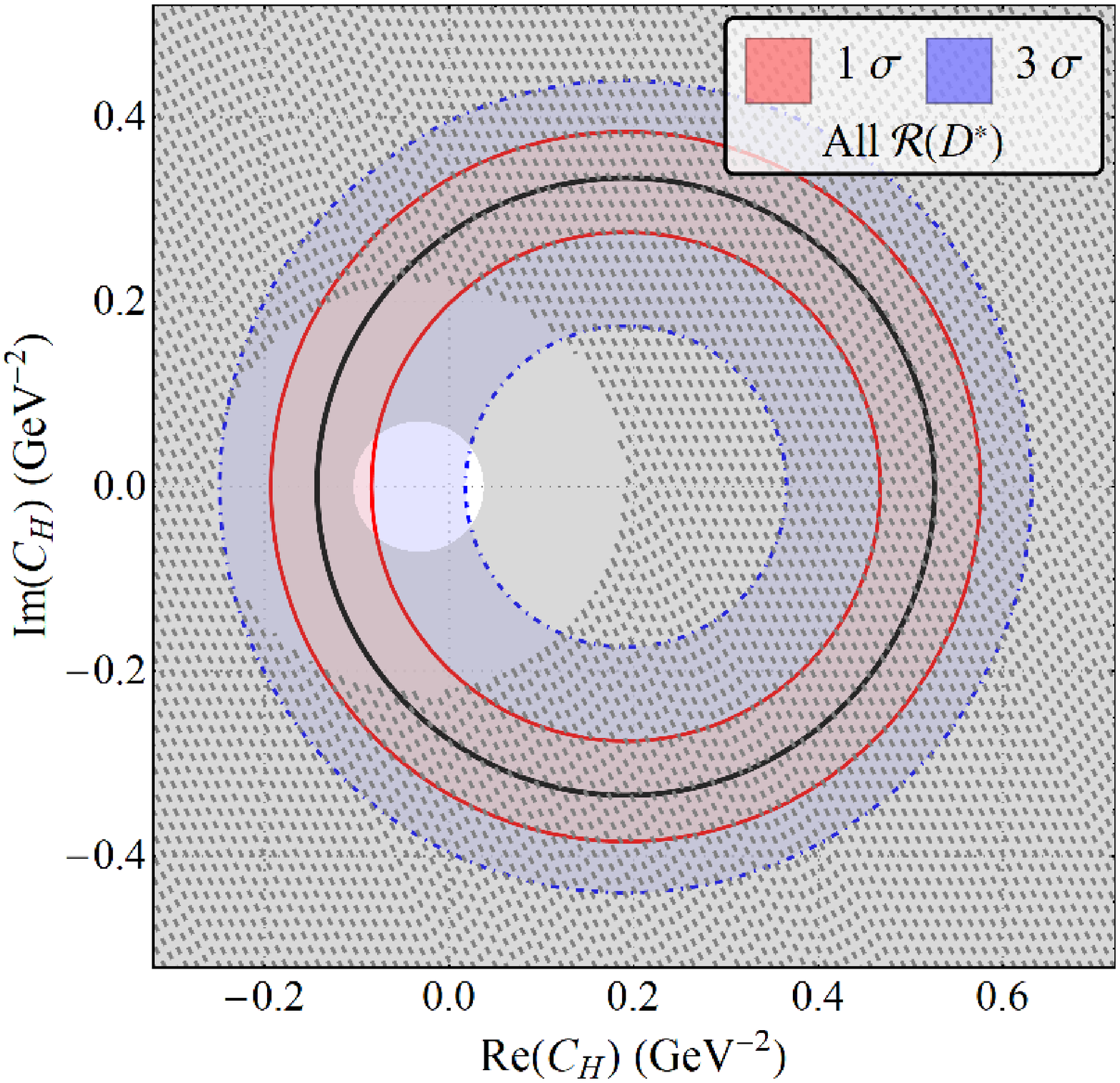}}
 		\caption{\small Fit results in terms of the fixed $\Delta \chi^2$ contours representing $1\sigma$ (red, solid) and $3\sigma$ (blue, dot-dashed) confidence levels respectively, in the Re($C_H$) and Im($C_H$) parameter-space for the case with $O_{S_1}$ only. The diagonally hatched region is ruled out from the $B_c$ life-time constraint and the gray-shaded region is disallowed by the constraint $\mathcal{B}(B_c \to \tau\nu) <10\%$.}
		\label{fig:rcHicH}
	\end{figure}

The results for our fits with a single vector and scalar type NP are displayed in fig.~\ref{fig:rcHicH} and table.~\ref{tab:res2}, assuming $c_H$ to be complex. 
	\begin{table}[H]
% 		\begin{center}
		\small
% 			\begin{ruledtabular}
				%\small
				\begin{tabular}{ccccccccc}
					& \multicolumn{2}{c}{Without $\mathcal{R}_{J/\psi}$} & \multicolumn{4}{c}{With $\mathcal{R}_{J/\psi}$} & \multicolumn{2}{c}{Fit Results}\\
					\cline{2-9}
					& & & \multicolumn{2}{c}{PQCD} & \multicolumn{2}{c}{LFCQ} & & \\
					\cline{4-7}
					Datasets & $\chi^2_{min}$ & $p$-value & $\chi^2_{min}$ & $p$-value & $\chi^2_{min}$ & $p$-value & Re($C_H$) & Im($C_H$) \\
					& /DoF  & (\%) & /DoF  & (\%) & /DoF  & (\%) & (GeV$^{-2}$)  & (GeV$^{-2}$) \\
					\hline
					All Data		& 9.22/8	& 23.72		& 11.86/9	&	15.76	& 12.38/9	&	13.51	& -0.031(8)		& 0.000(73) \\
					Belle			& 1.71/4	& 63.54		& 4.39/5	&	35.63	& 4.89/5	&	29.83	& -0.023(11)	& 0.000(87) \\
					Babar+LHCb	& 6.42/3	& 4.03		& 9.00/4	&	2.92 	& 9.54/4	&	2.29	& -0.042(11)	& 0.000(84) \\
					Babar+ Belle	& 6.71/6	& 24.31		& 9.35/7	&	15.48	& 9.87/7	&	13.03	& -0.030(8)		& 0.000(74) \\
					Belle + LHCb	& 4.70/6	& 45.41		& 7.37/7	&	28.82	& 7.88/7	&	24.72	& -0.025(11)	& 0.000(78) \\
					All $\mathcal{R}_{D^{*}}$	& 2.37/5	& 66.78		& 4.31/6 	&	50.53	& 4.99/6	&	41.67	& -	& - \\
					No $P_{\tau}(D^*)$	& 9.21/7	& 16.23	& 11.84/8	&	10.58	& 12.36/8	&	8.92	& -0.031(8)		& 0.000(72) \\
				\end{tabular}
 				\caption{\small Results of fits with different combinations of experimental data-points for the case with only $O_{S_1}$.} 
				\label{tab:res2}
% 			\end{ruledtabular}
% 		\end{center}
	\end{table}
\section{Models} 
The model independent scenarios described in the previous sections are further illustrated with the help of benchmark models in this section. Details about the model parameters and their relations with $C_{S_1}$ and $C_{V_1}$ can be found in~\cite{Biswas:2017vhc,Biswas:2018jun}.  
\subsection{Real $C_{V_1}$ and $C_{S_1}$: Non-Minimal universal extra dimension(NMUED)}
\begin{figure*}
\centering
\subfloat[$R_V$ vs. $R_f$]{
\includegraphics[width=4cm, height=4cm]{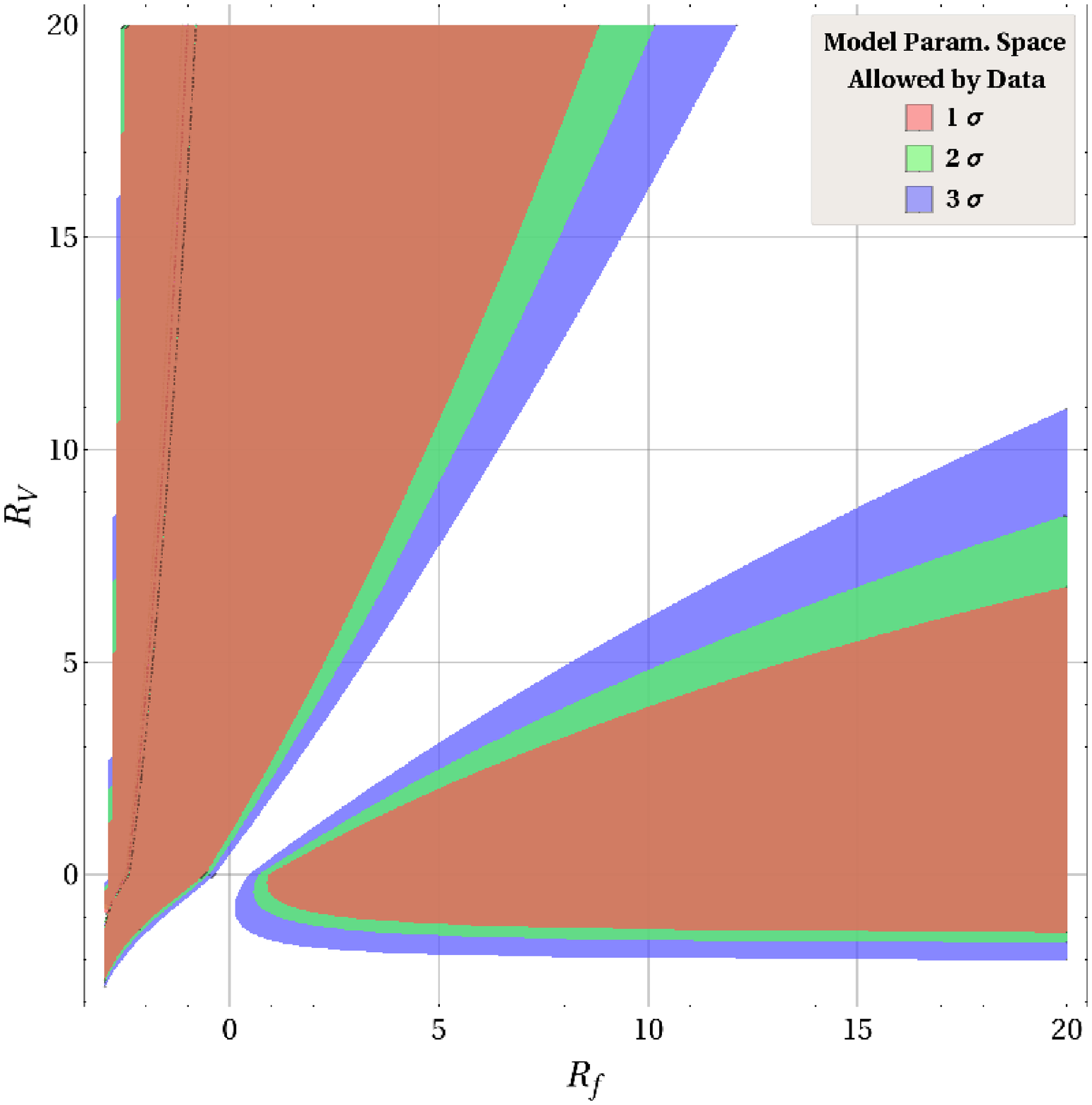}
 \label{fig:rfrvpltparam}}
\subfloat[$R_f$ vs. $R^{-1}$]{
\includegraphics[width=4cm, height=4cm]{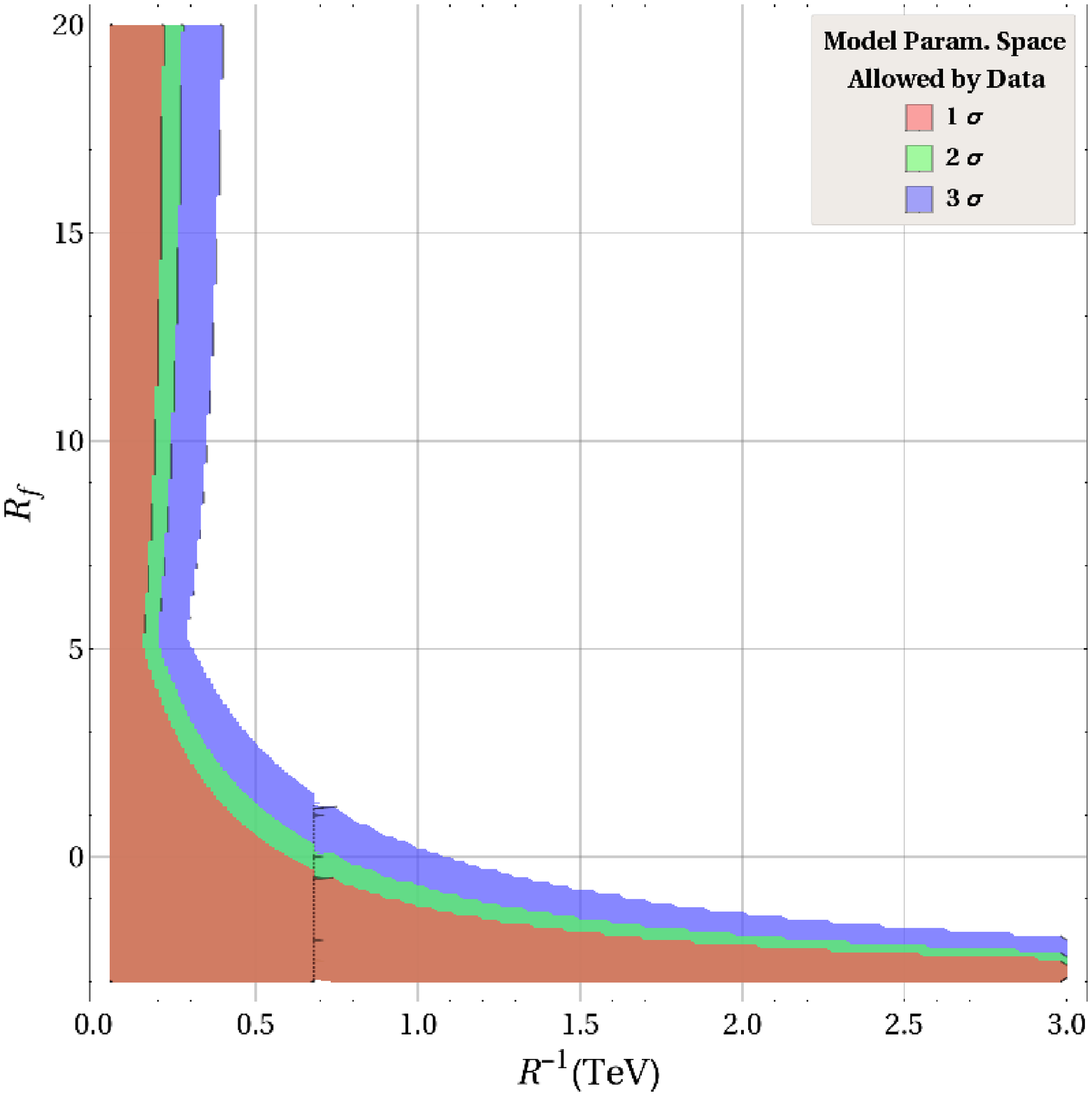}
 \label{fig:rrfpltparam}}
\subfloat[$R_V$ vs. $R^{-1}$]{
\includegraphics[width=4cm, height=4cm]{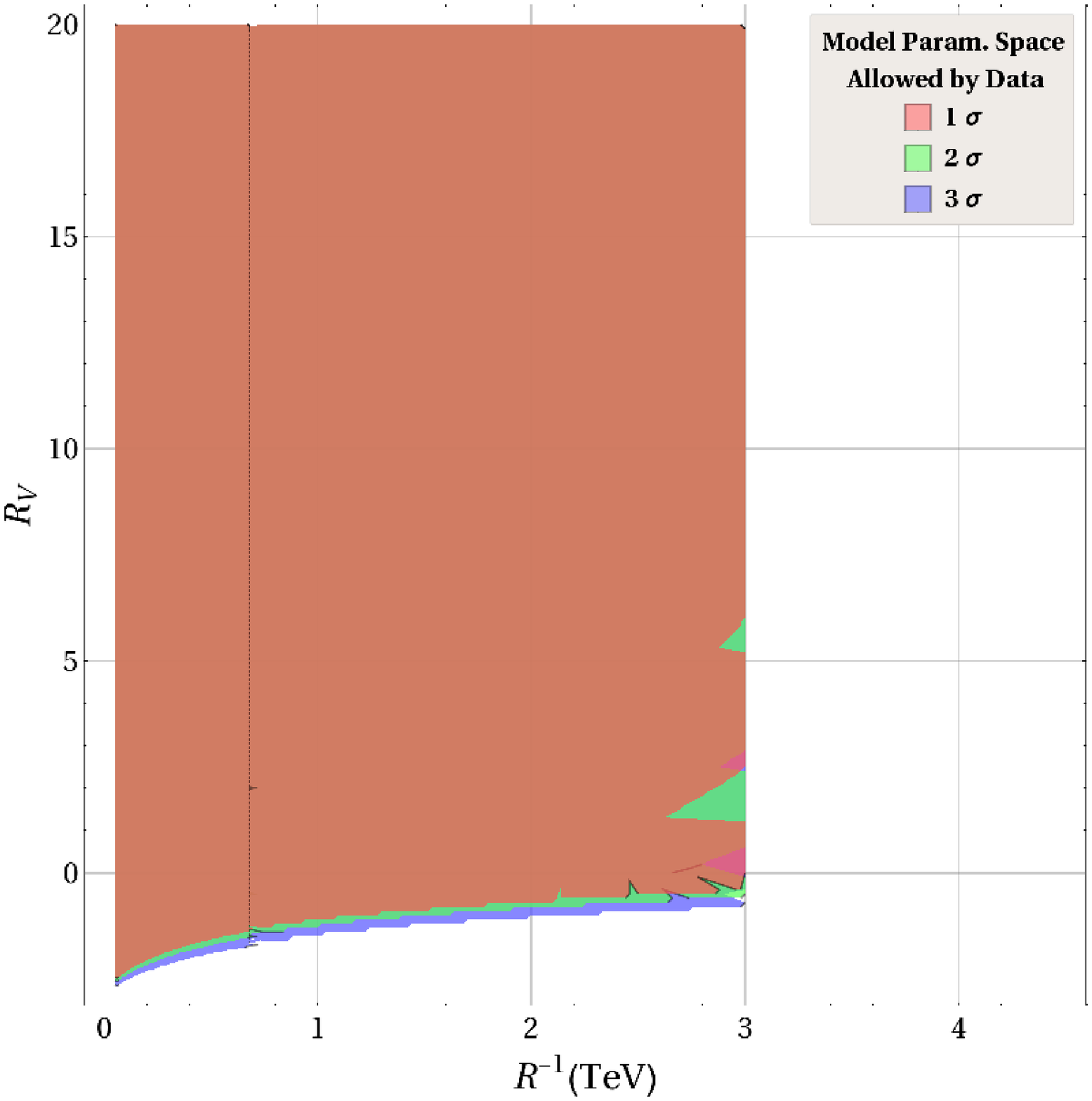}
 \label{fig:rrvpltparam}}
 \caption{Regions in the NMUED model parameter space, allowed by $C_W$ - $C^{\tau}_H$ fit of $\mathcal{R}(D^{(*)})$ data.}
\label{fig:modpar2}
\end{figure*}

\subsection{Real $C_{S_1}$: Goergi Michacek model (GM)}
	\begin{figure}[H]
		\centering
		\includegraphics[width=50mm,height=50mm]{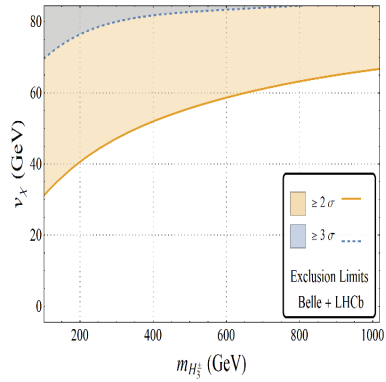}
		\caption{$v_{\chi}$ vs.$m_{H^{\pm}_3}$ parameter space excluded by all Belle and LHCb data at $2\sigma$ (orange, solid) and $3\sigma$ (blue, dashed) confidence levels. Regions above the lines are excluded.}
		\label{fig:excGM}
	\end{figure}

\subsection{Complex $C_{S_1}$: Leptoquark model (LQ)}
\begin{table*}[h!]
		\centering
		\small
			%\begin{ruledtabular}
%                     \resizebox{8cm}{!}{
			\begin{tabular}{|c|c|c|}\hline
		Data 		& ${\rm Re}\left(g^{33}_{2L}g^{23*}_{2R}\right)$ 	&  ${\rm Im}\left(g^{33}_{2L}g^{33*}_{2R}\right)$   \\
		 \hline
		All Data 						& $-0.250(64)$ 		& $0.0(6)$ \\
		Belle 							& $-0.186(90)$ 		& $0.0(7)$ \\
		Babar+LHCb 					& $-0.338(89)$ 		& $0.0(7)$ \\
		Babar + Belle 					& $-0.245(65)$ 		& $0.0(6)$ \\
		Belle + LHCb 					& $-0.198(88)$ 		& $0.0(6)$ \\
		No $P_{\tau }\left(D^*\right)$ 	& $-0.250(64)$ 		& $0.0(6)$ \\ \hline
			\end{tabular}
			%}
			\caption{Allowed values of the product of the couplings (both real and imaginary) of the chosen Leptoquark model involved with the Wilson coefficient $C_{S_1}^l$.} 
			\label{tab:LQ_Ch}
		%\end{ruledtabular}
	%\end{center}
\end{table*}

\end{document}